\documentclass[twocolumn,amsmath,amssymb,aps,preprintnumbers,showpacs,superscriptaddress]{revtex4}

\usepackage{bm}         
\usepackage{amssymb}
\usepackage{amsmath}
\usepackage{graphics}
\usepackage{epsfig}
\usepackage{color}
\usepackage{ulem}

\begin{document}

\title{Reducing vortex losses in superconducting microwave resonators with microsphere patterned antidot arrays}

\author{D.~Bothner}
\address{Physikalisches Institut and Center for Collective Quantum Phenomena in LISA$^+$, Universit\"{a}t T\"{u}bingen, Auf der Morgenstelle 14, D-72076 T\"{u}bingen, Germany}
\author{C.~Clauss}
\address{1. Physikalisches Institut, Universit\"{a}t Stuttgart, Pfaffenwaldring 57, D-70550 Stuttgart, Germany}
\author{E.~Koroknay}
\affiliation{Institut f\"{u}r Halbleiteroptik und Funktionelle Grenzfl\"{a}chen and Research Center SCoPE, Universit\"{a}t Stuttgart, Allmandring 3, D-70569 Stuttgart, Germany}
\author{M.~Kemmler}
\affiliation{Physikalisches Institut and Center for Collective Quantum Phenomena in LISA$^+$, Universit\"{a}t T\"{u}bingen, Auf der Morgenstelle 14, D-72076 T\"{u}bingen, Germany}
\author{T.~Gaber}
\affiliation{Physikalisches Institut and Center for Collective Quantum Phenomena in LISA$^+$, Universit\"{a}t T\"{u}bingen, Auf der Morgenstelle 14, D-72076 T\"{u}bingen, Germany}
\author{M.~Jetter}
\affiliation{Institut f\"{u}r Halbleiteroptik und Funktionelle Grenzfl\"{a}chen and Research Center SCoPE, Universit\"{a}t Stuttgart, Allmandring 3, D-70569 Stuttgart, Germany}
\author{M.~Scheffler}
\affiliation{1. Physikalisches Institut, Universit\"{a}t Stuttgart, Pfaffenwaldring 57, D-70550 Stuttgart, Germany}
\author{P.~Michler}
\affiliation{Institut f\"{u}r Halbleiteroptik und Funktionelle Grenzfl\"{a}chen and Research Center SCoPE, Universit\"{a}t Stuttgart, Allmandring 3, D-70569 Stuttgart, Germany}
\author{M.~Dressel}
\affiliation{1. Physikalisches Institut, Universit\"{a}t Stuttgart, Pfaffenwaldring 57, D-70550 Stuttgart, Germany}
\author{D.~Koelle}
\affiliation{Physikalisches Institut and Center for Collective Quantum Phenomena in LISA$^+$, Universit\"{a}t T\"{u}bingen, Auf der Morgenstelle 14, D-72076 T\"{u}bingen, Germany}
\author{R.~Kleiner}
\affiliation{Physikalisches Institut and Center for Collective Quantum Phenomena in LISA$^+$, Universit\"{a}t T\"{u}bingen, Auf der Morgenstelle 14, D-72076 T\"{u}bingen, Germany}
\date{\today}

\begin{abstract}
We experimentally investigate the vortex induced energy losses in niobium coplanar waveguide resonators with and without quasihexagonal arrays of nanoholes (antidots), where large-area antidot patterns have been fabricated using self-assembling microsphere lithography.
We perform transmission spectroscopy experiments around 6.25 and 12.5 GHz in magnetic field cooling and zero field cooling procedures with perpendicular magnetic fields up to $B=27$\,mT at a temperature $T=4.2$\,K.
We find that the introduction of antidot arrays into resonators reduces vortex induced losses by more than one order of magnitude.
\end{abstract}

\pacs{74.25.Qt, 74.25.Wx, 84.40.Dc, 03.67.Lx}

\maketitle

The importance of superconducting microwave circuitry devices has continuously grown during the last years.
In particular, coplanar microwave resonators have become very popular and are used in various fields such as circuit quantum electrodynamics \cite{Wallraff04, Hofheinz09, Niemczyk10}, quantum information processing \cite{DiCarlo09} and kinetic inductance particle detection \cite{Day03}.
As low energy losses are an essential requirement to these resonators, there are currently many efforts to identify and minimize the individual dissipation mechanisms \cite{Wang09, Macha10, Lindstroem09}.
Most recently it has been proposed to couple superconducting devices \cite{Rabl06, Imamoglu09, Verdu09, Henschel10, Bushev10} to trapped magnetic molecules, single electrons or ultra-cold atom clouds\cite{Fortagh05, Bushev08}, which requires superconducting low-loss circuitry to be operated in magnetic environments.
However, operating type-II superconducting resonators in magnetic fields leads to considerable energy dissipation due to Abrikosov vortex motion \cite{Song09}.
Recently some first approaches have been made to reduce the vortex associated losses under special experimental conditions.
Amongst others \cite{Schuster10, Song09a}, patterning the resonators with antidots, well-known and highly controllable pinning sites for Abrikosov vortices \cite{Fiory78, Woerdenweber04}, has shown promising results \cite{Bothner11}.
It was demonstrated that a few antidots with diameters on the micron scale, strategically positioned at the resonator edges, are able to significantly reduce vortex associated losses in zero field cooling (zfc) experiments with magnetic fields of $\sim$10 Gauss.
In addition we found that the losses decreased with increasing number of antidots.
Thus, in order to expand the magnetic field range for the operation of the resonators it is necessary to increase the antidot density and at the same time reduce the antidot diameter. 
In the past there have been various approaches based on self-assembling techniques \cite{Welp02, Vinckx06, Eisenmenger07}, as they allow to cover the whole substrate area with a large amount of tiny antidots.
Yet, for high-frequency applications these techniques cannot be easily applied, as some of them depend on the substrate material \cite{Jessensky98, Welp02} itself and most of them change the substrate properties during the fabrication process significantly.
In this letter we propose and demonstrate a different fabrication technique suitable for large area antidot lattices in superconducting microwave resonators, which to the best of our knowledge is novel in the context of superconducting thin films and vortex pinning.
We use self-assembled polystyrene colloids as a quasi-hexagonal array of microlenses which can be used for the patterning of antidots by optical lithography \cite{Wu08}.
Such patterned samples show vortex associated microwave losses in perpendicular magnetic fields $B\leq27$\,mT, which are more than one order of magnitude smaller than for reference samples without antidots.
The presented results may also be relevant for other superconducting thin film microelectronic devices, \textit{e.g.} kinetic inductance detectors, filters or quantum interference devices, when operated in external magnetic fields.
The fabrication process starts with dc magnetron sputtering of a $\sim150$\,nm thick Nb layer onto a $2^{\prime\prime}$ r-cut sapphire wafer.
These Nb films have a critical temperature $T_c\approx 9\,$K and a residual resistance ratio of $R_{300\rm{\,K}}/R_{10\rm{\,K}}\approx 6$.
After cutting the wafer into single chips, lithography was performed. 
We covered the chips with photoresist and transfered a monolayer of water suspended polystyrene microspheres onto them in a Langmuir-Blodgett deposition process.
The microspheres have an average diameter of $770\,$nm ($\pm25\,$nm) and act as microlenses for UV-light, leading to a focused energy density directly below each sphere and thus after development to a quasihexagonally arranged hole pattern.
With reactive ion etching ($\rm SF_6$) the hole array was transferred into the Nb film.
Electric transport measurements on test structures with antidots, which were performed close to $T_c$, showed pronounced matching effects.
From these measurements we estimate the antidot density to $n_p\approx 1.05\,\mu$m$^{-2}$.
A detailed discussion of these results together with a deeper analysis of the properties of the self organized quasi-hexagonal antidot lattice will be given elsewhere.
\begin{figure}[h]
\includegraphics[scale=0.43]{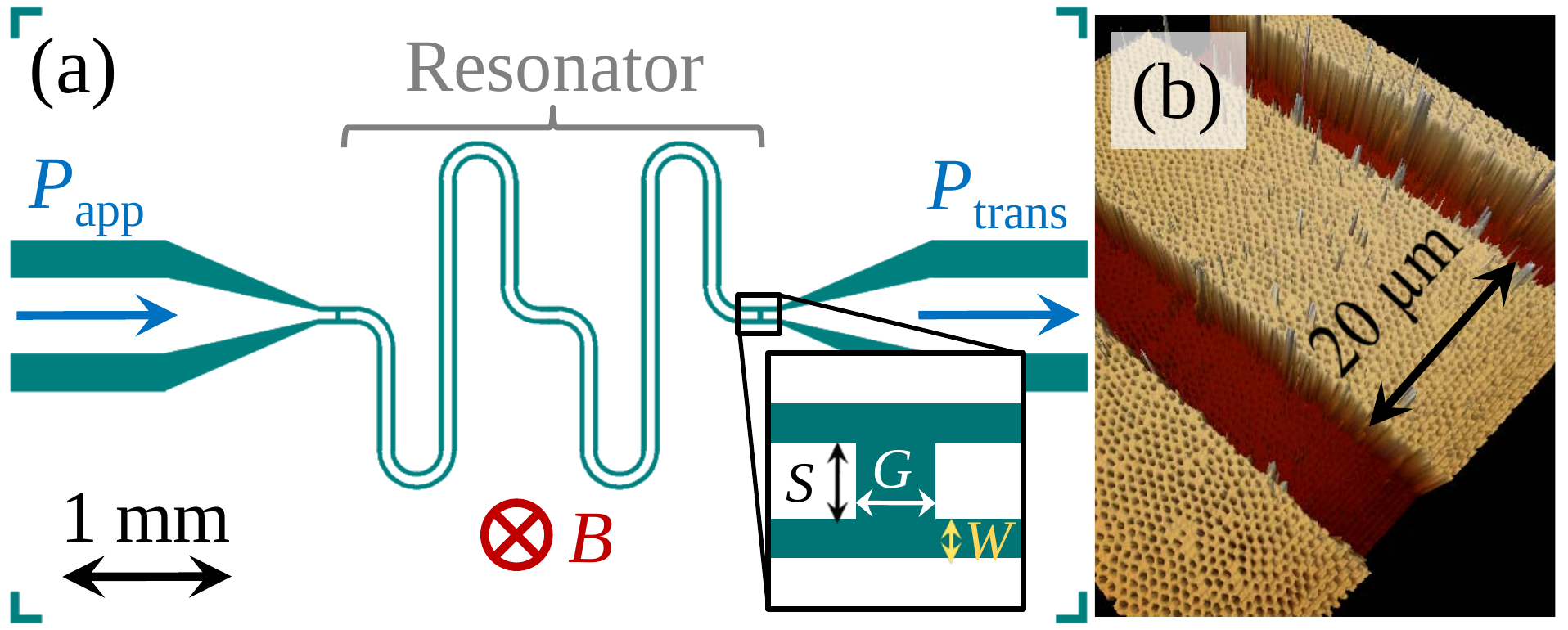}
\caption{(Color online) (a) Layout of a $7\times4$\,mm$^2$ chip with a capacitively coupled 6.2 GHz transmission line resonator (parameter set 2). (b) AFM image of a part of a resonator perforated with antidots (parameter set 1).
}
\label{fig:Graph1}
\end{figure}
Into the perforated as well as plain Nb films we patterned half wavelength transmission line resonators with a resonance frequency $f_{\rm{res}}\approx 6.25$\,GHz using standard optical lithography and reactive ion etching ($\rm SF_6$).
The resonators are capacitively and symmetrically coupled to feed lines via gaps at both ends.
For this study we fabricated two types of resonators with the same general layout and length, but with different widths of the center conductor ($S_1=20$\,$\mu$m, $S_2=60$\,$\mu$m), of the gap between center conductor and ground planes ($W_1=8$\,$\mu$m, $W_2=25$\,$\mu$m) and of the coupling gaps ($G_1=10$\,$\mu$m, $G_2=30$\,$\mu$m).
Both designs result in a characteristic impedance of the transmission line $Z_0\approx50$\,$\Omega$.
Figure \ref{fig:Graph1} shows (a) a sketch of the resonator layout (parameter set 2) and (b) a partial AFM image of a resonator with antidots (parameter set 1).
For the characterization of the resonators we mounted each of them into brass boxes and contacted them with silver paste to sub-miniature A (SMA) stripline connectors (signal line/center conductor) and the box (ground planes).
We determined the frequency dependent transmitted power $P_{\rm{trans}}$ with a spectrum analyzer for different values of perpendicularly applied magnetic field $|B|\leq 27$\,mT at a temperature $T=4.2\,$K.
No attenuators or amplifiers were used in the measurements, but due to cable and connector losses we estimate the effective power at the resonator input to be at least $10\,$dB lower than at the generator output, to which all values of the variable applied power $P_{\rm{app}}$ refer.
We fit each resonance curve $P_{\rm{trans}}(f)$ with a Lorentzian, from which we extract the resonance frequency $f_{\rm{res}}(B)$ and the width of the resonance at half maximum $\Delta f(B)$.
This allows us to calculate the quality factor $Q(B)=f_{\rm{res}}(B)/\Delta f(B)$ and alternatively, the energy loss $1/Q(B)$.
In general, the overall losses $1/Q(B)$ are the result of several different (field independent and field dependent) loss mechanisms such as dielectric, resistive or radiative losses.
The quantity $1/Q_v(B)=1/Q(B)-1/Q(0)$ allows us to extract the field dependent contribution, which we associate with Abrikosov vortices, cf. also \cite{Song09, Bothner11}.
In a first measurement series we performed field cooling (fc) experiments, \textit{i.e.} we apply each magnetic field value at a temperature above $T_c$ and subsequently cool down the resonator below $T_c$.
With this procedure the vortices are expected to be nearly homogeneously distributed within the film.
\begin{figure}[h]
\includegraphics[scale=0.4]{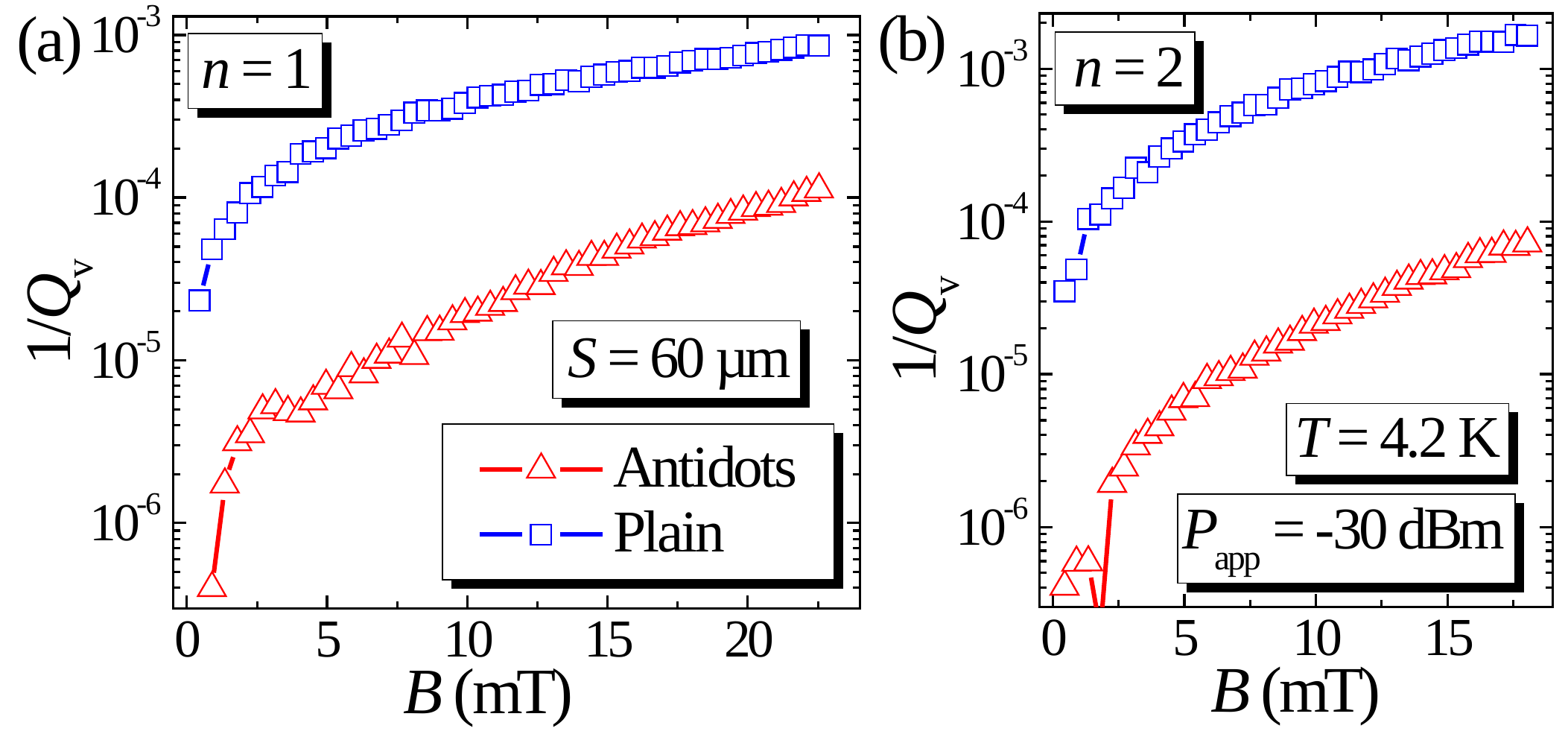}
\caption{(Color online) Comparison of the vortex associated energy losses $1/Q_v(B)$ in fc measurements for a perforated resonator (triangles) and a plain resonator (squares) with identical design parameters: (a) fundamental $n=1$ (corresponding to $f^{\textrm{Plain}}_{\textrm{res}}(0) = 6.27\,$GHz and $f^{\textrm{Antidots}}_{\textrm{res}}(0) = 6.25\,$GHz) and (b) second harmonic $n=2$ ($f^{\textrm{Plain}}_{\textrm{res}}(0) = 12.53\,$GHz and $f^{\textrm{Antidots}}_{\textrm{res}}(0) = 12.47\,$GHz).}
\label{fig:Graph2}
\end{figure}
Figure~\ref{fig:Graph2}(a) compares $1/Q_v(B)$ for the fundamental mode $n=1$ of two resonators with identical parameters (set 2): one with and one without antidots.
The resonator with antidots shows more than one order of magnitude lower losses (maximum factor for the fundamental mode is 40) than its plain counterpart over the whole investigated field range.
We attribute this enhanced performance to an effective trapping and pinning of vortices by the antidots.
We also performed measurements for higher resonator modes, \textit{i.e.} $n=2,3$, which show similar results. 
Note, at higher frequencies ($n>1$) and magnetic fields the cavity resonance is significantly suppressed, especially in plain resonators, such that for some $B>B_n$ the resonance curve cannot be reliably isolated from parasitic resonances anymore.
For the first harmonic, \textit{i.e.} $n=2$, vortex losses $1/Q_v(B)$ could be extracted for $B\leq 18$\,mT. Results are depicted in Fig.~\ref{fig:Graph2} (b), showing a comparable improvement of the performance by the antidots as for $n=1$.
%
%
We find a similar behavior for resonators with parameter set 1.

\begin{figure}[h]
\includegraphics[scale=0.4]{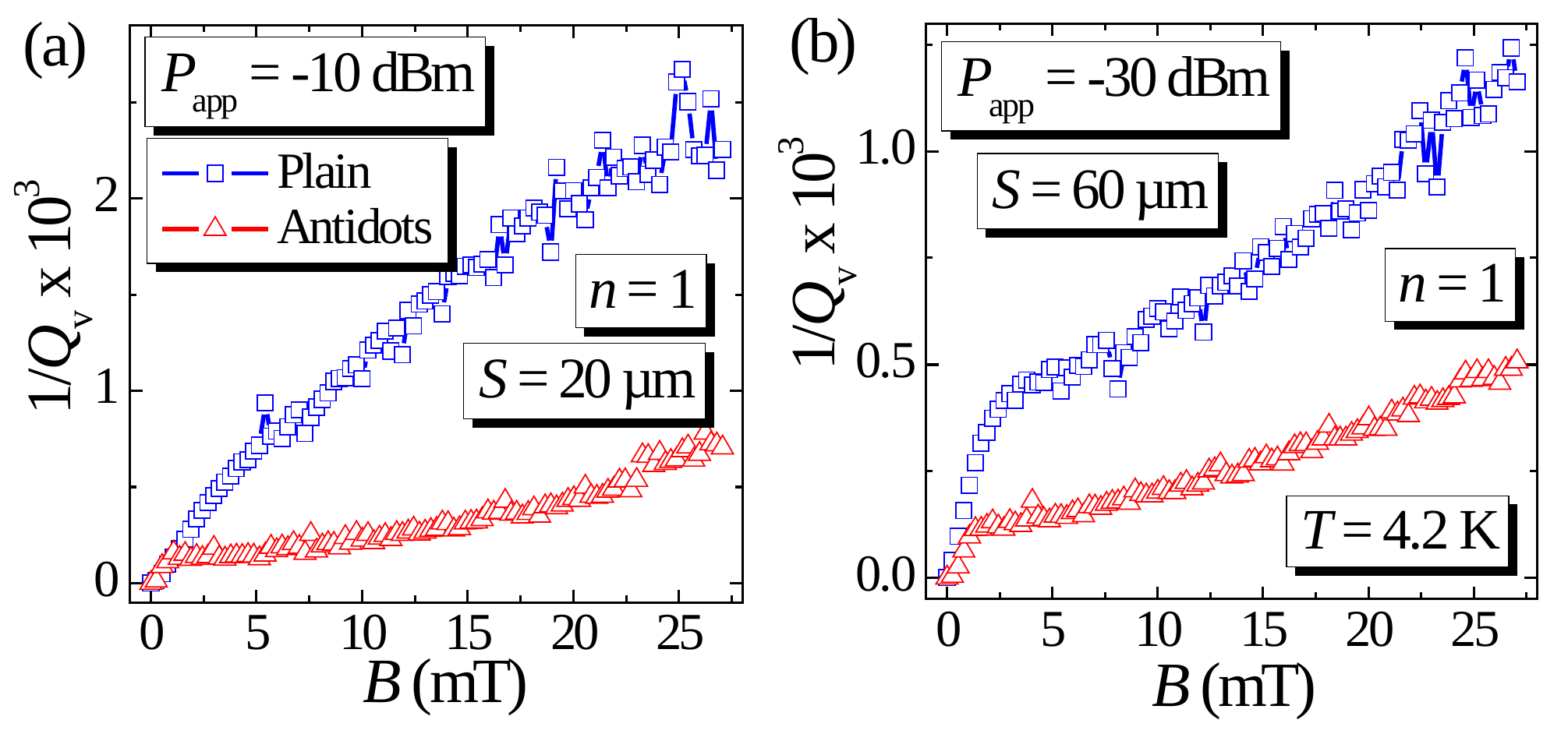}
\caption{(Color online) Vortex associated energy losses $1/Q_v(B)$ for zfc measurements of two resonators with (a) parameter set 1 and (b) parameter set 2. One resonator of each parameter set was patterned with antidots (red triangles), one was not (blue squares).}
\label{fig:Graph3}
\end{figure}
In a second set of experiments resonators were cooled down to $T=4.2\,$K in $B=0$ before applying a magnetic field.
Here, the vortices are expected to penetrate the superconductor with a Bean-like flux profile \cite{Brandt93}, which leads to an \textit{enhanced} flux density close to the edges of the Nb structures.
Figure~\ref{fig:Graph3} shows the vortex associated losses $1/Q_v(B)$ of (a) two resonators with parameter set 1 and (b) two resonators with parameter set 2.
Within each parameter set we compare a resonator with antidots to a plain one.
We find higher losses $1/Q_v(B)$ than in the corresponding fc measurements.
This result, which at first glance is somehow counterintuitive, since the number of vortices within the superconductor is smaller for the zfc than for the fc case, has already been reported before \cite{Lahl03, Ghigo07}.
It can be explained by the inhomogeneous flux distribution for the zfc situation which has its maximum near the Nb film edges. 
Since the microwave current density is also higher at the edges, this leads to increased losses $1/Q_v(B)$ when compared to the field-cooled case.
In the zfc case, for both parameter sets the resonator with antidots shows significantly reduced losses $1/Q_v(B)$ over the whole field range (up to a factor of about 6).
Yet, the absolute values of $1/Q_v(B)$ are higher for resonators with parameter set 1 than those of parameter set 2.
There are two possible mechanisms, which may explain this observed behavior.
First, as parameter set 1 corresponds to smaller gaps between the center conductor and the ground planes, the flux focusing and hence the effective flux density seen by the resonator is higher.
Second, in the zfc measurements, we sometimes observe sudden jumps in the transmitted power corresponding to a small shift of the resonance curve which are an indication for the appearance of flux avalanches \cite{Ghigo07} (also represented in the noise of $1/Q_v(B)$).
Those flux avalanches reduce the flux density at the conductor edges by transporting flux towards the center of the conductor, where the microwave current density is considerably smaller. 
Obviously, the flux can be transported farther away from the edges for the case of a wider center conductor.  
Therefore, flux avalanches in resonators with parameter set 2 may result in a stronger flux reduction at the edges of the center conductor than in resonators with parameter set 1. 
%

%

%
In conclusion, we have used a novel patterning technique for the fabrication of large area periodic micron-sized antidot arrays in superconducting thin film microwave devices based on self-assembled microspheres to significantly improve the device performance in magnetic fields.
We have demonstrated, that energy losses in superconducting microwave resonators due to the presence of Abrikosov vortices can be reduced by more than one order of magnitude with quasihexagonal antidot arrays.
We have shown this result to be valid for magnetic fields up to at least $B\approx 27\,$mT for fc procedures as well as for zfc measurements.
This work has been supported by the Deutsche Forschungsgemeinschaft via SFB/TRR 21 and by the European Research Council via SOCATHES.
D. B. acknowledges support by the Evangelisches Studienwerk Villigst e.V., M. K. acknowledges support by the Carl-Zeiss Stiftung.

\end{document}